\documentclass[apjl]{emulateapj}

\usepackage{graphicx}

\slugcomment{The Astrophysical Journal Letters, in press}
\shorttitle{A collisional origin for the Leo ring}
\shortauthors{Michel-Dansac et al.}

\begin{document}

\newcommand{\hi}{{\sc H\,i}}

\title{A collisional origin for the Leo ring\footnote{Based on
    observations obtained with MegaPrime/MegaCam, a joint project of
    CFHT and CEA/DAPNIA, at the Canada--France--Hawaii Telescope (CFHT)
    which is operated by the National Research Council (NRC) of
    Canada, the Institute National des Sciences de l'Univers of the
    Centre National de la Recherche Scientifique of France, and the
    University of Hawaii.}}








\author{Leo Michel-Dansac\altaffilmark{1}, 
  Pierre-Alain Duc\altaffilmark{2}, Frederic Bournaud\altaffilmark{2},
  Jean-Charles Cuillandre\altaffilmark{3}, Eric
  Emsellem\altaffilmark{4,1}, Tom
  Oosterloo\altaffilmark{5}, Raffaella Morganti\altaffilmark{5}, Paolo
  Serra\altaffilmark{5}, and Rodrigo Ibata\altaffilmark{6}}

\altaffiltext{1}{Centre de Recherche Astrophysique de Lyon, Universit\'e de Lyon,
  Universit\'e Lyon 1, Observatoire de Lyon, Ecole Normale Sup\'erieure de
  Lyon, CNRS, UMR 5574, 9 avenue Charles Andr\'e, 69561
  Saint-Genis-Laval cedex, France; leo@obs.univ-lyon1.fr}

\altaffiltext{2}{Laboratoire AIM, CEA/Irfu, CNRS, Universit\'e Paris Diderot,
  SAp, 91191 Gif-sur-Yvette cedex, France}

\altaffiltext{3}{Canada--France--Hawaii Telescope, P.O. Box 1597, Kamuela, HI 96743, USA}

\altaffiltext{4}{European Southern Observatory, Karl-Schwarzschild-Str. 2, 85748 Garching bei Muenchen, Germany}

\altaffiltext{5}{ASTRON, Netherlands Institute for Radio Astronomy, Postbus 2, 7990 AA Dwingeloo, The Netherlands}

\altaffiltext{6}{Observatoire Astronomique de Strasbourg (UMR7550), 11, rue de
  l'Universit\'e, 67000 Strasbourg, France}

\begin{abstract}

Extended \hi\ structures around galaxies are of prime importance to
probe galaxy formation scenarios. The giant \hi\ ring in the Leo group
is one of the largest and most intriguing \hi\ structures in the
nearby universe. Whether it consists of primordial gas, as suggested
by the apparent absence of any optical counterpart and the absence of
an obvious physical connection to nearby galaxies, or of gas expelled
from a galaxy in a collision is actively debated.  We present deep
wide field-of-view optical images of the ring region obtained with
MegaCam on the CFHT. They reveal optical counterparts to several
\hi\ and UV condensations along the ring, in the $g'$, $r'$, and $i'$ bands,
which likely correspond to stellar associations formed within the
gaseous ring.  Analyzing the spectral energy distribution of one of
these star-forming regions, we found it to be typical for a
star-forming region in pre-enriched tidal debris. We then use
simulations to test the hypothesis that the Leo ring results from a
head-on collision between Leo group members NGC~3384 and M96.
According to our model which is able to explain, at least
qualitatively, the main observational properties of the system, the
Leo ring is consistent with being a collisional ring. It is thus
likely another example of extended intergalactic gas made-up of
pre-enriched collisional debris.

\end{abstract}

\keywords{galaxies: evolution -- galaxies: groups: individual (Leo
  group) -- galaxies: individual (NGC 3384, M96) -- galaxies:
  interactions}

\section{Introduction}

The quest for primordial gas clouds that have never been involved in
star-forming episodes in the local universe has motivated extensive
\hi\ surveys (e.g., HIPASS, \citealp{hipass}; and ALFALFA,
\citealp{alfalfa}) but has yet been rather unsuccessful. Among the
putative candidates are the so-called dark galaxies
\citep[e.g.,][]{Davies04}, which are however likely tidal debris from
galaxy interactions \citep{Duc08}. Isolated \hi\ clouds around
early-type galaxies \citep[ETGs;][]{Morganti06,Oosterloo07,Serra10}
could either be accreting clouds or the collisional remnants of the
violent merger events that are usually invoked for the formation of
ETGs. Investigating the stellar populations associated or not with these
clouds could discriminate these formation scenarios.

The Leo ring, in the Leo group of galaxies, is one of the most
spectacular and mysterious intergalactic \hi\ structures known in the
nearby universe
\citep{Schneider83,Schneider85,Schneider89b,Stierwalt09}.  It has a
ring-like shape of diameter $\sim 200$~kpc, quite asymmetric and
somewhat clumpy, apparently centered on the NGC~3384/M105 galaxy
pair. Some radial filamentary structures are observed, in particular a
bridge connecting the ring to the spiral galaxy M96.

Numerous attempts were done to find counterparts to the \hi\ emission
at other wavelengths, which remained unsuccessful in the optical and
H$\alpha$ regime. This had suggested that the ring could consist of
primordial gas. The only relatively bright optical counterparts
correspond to \hi-rich dwarf galaxies in the group \citep{Stierwalt09}
whose velocities are discrepant from the ring velocity field (e.g., HGC
201970)\footnote{One exception is HGC 202027, whose Sloan Digital Sky
  Survey (SDSS) redshift is
  close to that of the \hi\ ring, but is not associated with an
  \hi\ clump peak of the ring. It is most likely a pre-existing group
  member unrelated to the formation of the \hi\ ring.}.

Lastly, \citet{Thilker09} reported a UV counterpart to some
\hi\ clumps along the ring from \textit{Galaxy Evolution Explorer (GALEX)} observations. The UV knots with
the highest far-UV to near-UV flux ratios are good candidates for
being star-forming regions associated with the ring (and not background
galaxies). The inferred star formation rate is very low but consistent
with the \hi\ column densities below $2 \times
10^{20}$~cm$^{-2}$. Based on the FUV--NUV color and upper limits on the
FUV--r color, \citet{Thilker09} estimated a very low gas metallicity,
and concluded that the \hi\ ring is likely primordial.
  
\begin{figure*}[th]
  \begin{center}
  \includegraphics[width=0.9\textwidth]{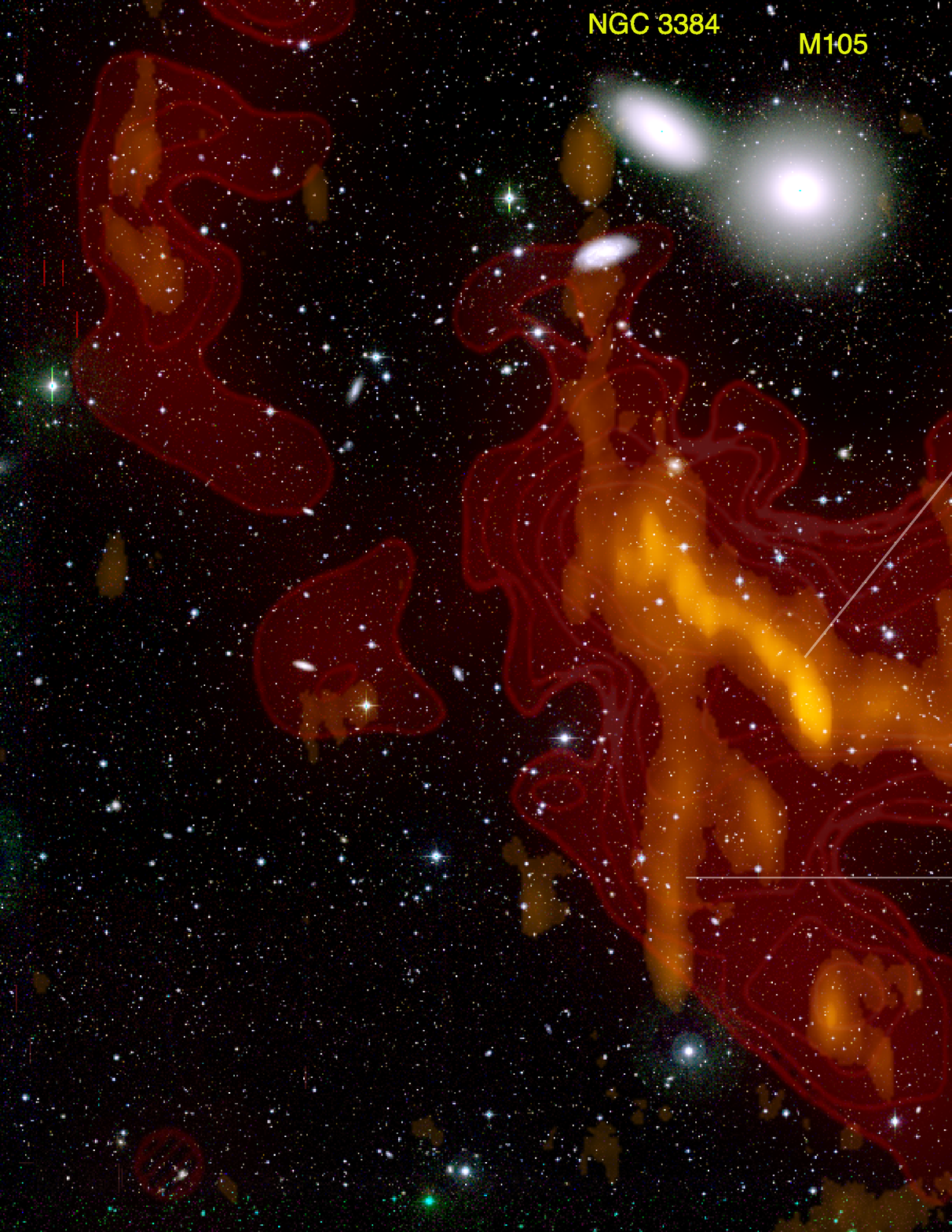}
  \caption{Leo ring. Left: distribution of the \hi\ gas, as
    mapped at Arecibo (reddish contours; from
    \citealt{Schneider89}), the WSRT (in brown; T. A. Oosterloo et al., in
    preparation), superimposed on a deep optical image obtained at the
    CFHT.  Right: zoom on several UV condensations
    \citep{Thilker09}. The \textit{GALEX} FUV (white contours) and WRST
    \hi\ emission (brown contours) are superimposed on the optical true
    color image made from $g'$, $r'$, and $i'$ bands data. The apertures
    used by \cite{Thilker09} are shown in green. }
  \end{center}
\label{fig:megacam}
\end{figure*}

More recently, \cite{Bot09} reported a marginal detection of dust
emission at 8 and 24 $\mu$m toward the densest \hi\ condensation. If
real, it implies that the metallicity of the \hi\ ring is higher than
the $Z_{\odot}$/20 estimated by \cite{Thilker09}. This would support
scenarios where the gas has a galactic origin. Indeed, clouds expelled
during galaxy interaction have been previously processed in the parent
galaxies and are thus relatively metal-rich. \cite{Bekki05b} proposed
that the disruption of a low surface brightness galaxy in the
gravitational potential of a group forms gaseous arcs and rings with
column density in agreement with observations, without providing a
specific model for the Leo ring, though.  Collisional rings are a more
classical example of such debris
\citep{Appleton96,Bekki98,Bournaud03a}: a head-on galaxy collision
disrupts a gaseous galactic disk into an expanding pre-enriched ring
of gas, whose star-forming activity depends on the remaining gas
density.

In this Letter, we present new optical images of the system and report
the detection of counterparts in the $g'$, $r'$, and $i'$ bands (Section
2). We present a numerical model of the Leo ring as the result of a
head-on collision in the Leo group (Section 3) and discuss this
hypothesis in Section 4.

\section{Observations of an optical counterpart to the Leo ring}

\begin{table}[ht]
  \begin{center}
    \caption{Photometry of UV knot 2E}
    \begin{tabular}{ccccc}
      \hline
      \hline
      FUV\tablenotemark{a} & NUV\tablenotemark{a} & $g'$ & $r'$ & $i'$\\
      \hline
      \multicolumn{5}{c}{ {\scriptsize AB mag corrected for Galactic
          extinction}} \\
      \hline
      {\scriptsize $21.6 \pm 0.15$} & {\scriptsize 21.4 $\pm$  0.2} &
      {\scriptsize 21.0 $\pm$  0.04} & {\scriptsize 20.8 $\pm$  0.04} & {\scriptsize 21.1 $\pm$  0.06}\\
      \hline
    \end{tabular}
    \footnotetext{Our photometric measurement resulted in larger
      error bars than those listed in \citet{Thilker09}, and fluxes
      closer to the values listed in the \textit{GALEX} online catalog. Aperture
      corrections and differences in the estimation of the background
      level might explain this.}
    \label{tab:fluxes}
  \end{center}
\end{table}

Deep optical observations of the Leo group of galaxies were obtained
in 2009 January, November, and December with the MegaCam camera on the
CFHT, as part of a survey of \hi-rich ETGs from the
ATLAS$^{\rm 3D}$ project\footnote{http://purl.org/atlas3d}
\citep{Serra09b}.

The whole \hi\ structure was covered with two MegaCam pointings of
1~$\deg^2$ each. Each pointing consisted of individual exposures,
offset with respect to each other by typically 10 arcmin. This allowed
to create a master sky which was subtracted before recombining the
images. Instrumental artifacts such as diffuse light could thus be
removed down to a surface brightness limit of 28.2 mag~arcsec$^{-2}$
in the $g'$ band. The total exposure times on the southern field (which
contains the densest \hi\ condensations) were 1764~s in the $g'$ band,
2760~s in the $r'$ band, and 714~s in the $i'$ band. Observations of the
northern field were shallower: 882~s, 1380~s, and 71~s in $g'$, $r'$, and
$i'$, respectively. Weather conditions were photometric. The data were
processed using the Elixir-LSB software originally developed as part
of the Next Generation Virgo Cluster Survey (L. Ferrarese et al. 2010,
in preparation).

A true color image field is presented in Figure~\ref{fig:megacam}
together with close-ups toward three UV-detected
\hi\ condensations. New \hi\ maps obtained with the Westerbork
Synthesis Radio Telescope (WSRT), to be presented in T. A. Oosterloo
et al. (2010, in
preparation) are superimposed on the optical images.

At the depth of our observations, no diffuse stellar emission is found
along the \hi\ ring.  However, several faint, very compact, blue knots
are detected for the first time (in particular clump 2E in
Figure~\ref{fig:megacam}) toward the UV knots. This seems to confirm
that the \textit{GALEX} detections are not spurious. If as claimed by
\cite{Thilker09} they are physically linked to the \hi\ condensations,
they most likely correspond to OB associations formed in situ within
the ring.
 
\begin{figure}
  \includegraphics[width=\columnwidth]{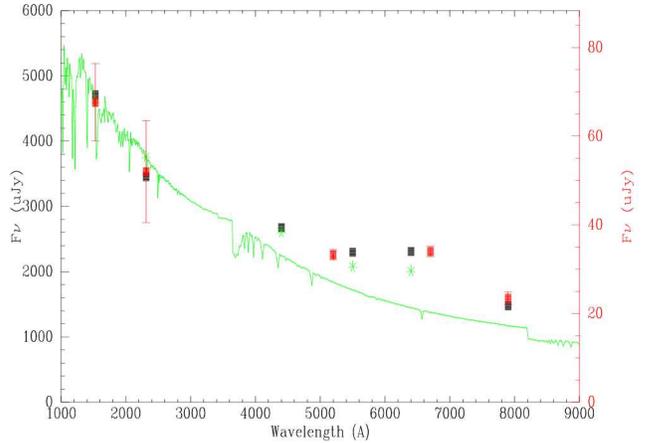}
  \caption{SED of a typical star-forming region in a collisional
    ring. The black filled squares correspond to NGC 5291N, the most
    luminous optical condensation in the collisional ring of NGC 5291
    \citep{Duc98b}, assuming an optical extinction of
    $E(B-V)=0.27$~mag, as constrained by stellar synthesis modeling
    \citep{Boquien10}. The best-fitting model shown in green (curve:
    stellar emission, symbols: total emission including the line
    emission) corresponds to an instantaneous, 5~Myr old starburst
    with an exponential decay timescale of 1~Myr and half-solar
    metallicity as measured in the {\sc H\,ii} regions of
    NGC~5291. The \citet{Calzetti00} extinction curves were used. The
    same model matches the SED of clump 2E in the Leo ring, shown with
    red squares, assuming an optical extinction of
    $E(B-V)=0.22$~mag. The data are shown after normalizing the same
    $R$-band flux: the scale on the left axis corresponds to NGC
    5291N, and on the right axis to Leo clump 2E.}
  \label{fig:SED}
\end{figure}

The multi-band aperture photometry on the UV clumps proved to be
challenging: indeed, with an average seeing of 0.9 arcsec, our optical
observations resolve the \textit{GALEX} UV knots into several optical
point-like sources displaying a variety of colors and thus possibly of
origin (see in particular clumps 1 and 2 in
Figure~\ref{fig:megacam}). Clump 2E avoids these problems: the \textit{GALEX}
and optical emission coincide spatially and the MegaCam sources show a
uniform blue color. We carried out a detailed photometric analysis on
this condensation. Our measurements in the NUV and FUV \textit{GALEX} bands,
and $g'$, $r'$, and $i'$ MegaCam bands are presented in
Table~\ref{tab:fluxes}. They were obtained by using the {\sc polyphot}
task of the IRAF package, with an aperture chosen to follow the
external white contours shown in Figure~\ref{fig:megacam}. Local sky
measurements were estimated manually to avoid nearby
background/foreground sources.

The UV to optical spectral energy distribution (SED) of clump 2E is
shown in Figure~\ref{fig:SED}. We compare it to the SED of NGC 5291N,
a luminous star-forming region in NGC 5291, which harbors an extended
ring of confirmed collisional debris \citep{Bournaud07} with a
morphology recalling that of the Leo ring. Accounting for a different
flux and a difference in extinction of 0.2~mag in $A_V$, the two SEDs
are strikingly similar (Figure~\ref{fig:SED}). The metallicity of NGC
5291N, as estimated from a measure of the oxygen abundance in its
ionized gas, is moderately high: 12+log(O/H)=8.4 \citep{Duc98b},
suggesting that this could also be the case for clump 2E. As recently
shown by \cite{Boquien10}, the SED of NGC 5291N, and thus of Leo clump
2E (see also Figure~\ref{fig:SED}), is well fit by a model depicting
an instantaneous starburst of age of about 5 Myr, within dust obscured
pre-enriched gaseous material.  Therefore, assuming that dust is
present in the Leo \hi\ structure, as suggested by the marginal
\textit{Spitzer} detections reported by \cite{Bot09}, the SED of clump 2E is
consistent with the hypothesis that it was formed within pre-enriched
material\footnote{The large error bars in the UV preclude a direct fit
  of the SED of this specific object with spectro-photometric
  models.}.

\begin{figure*}


  \begin{center}
  \begin{minipage}{6.1cm}
    \includegraphics[width=3.cm]{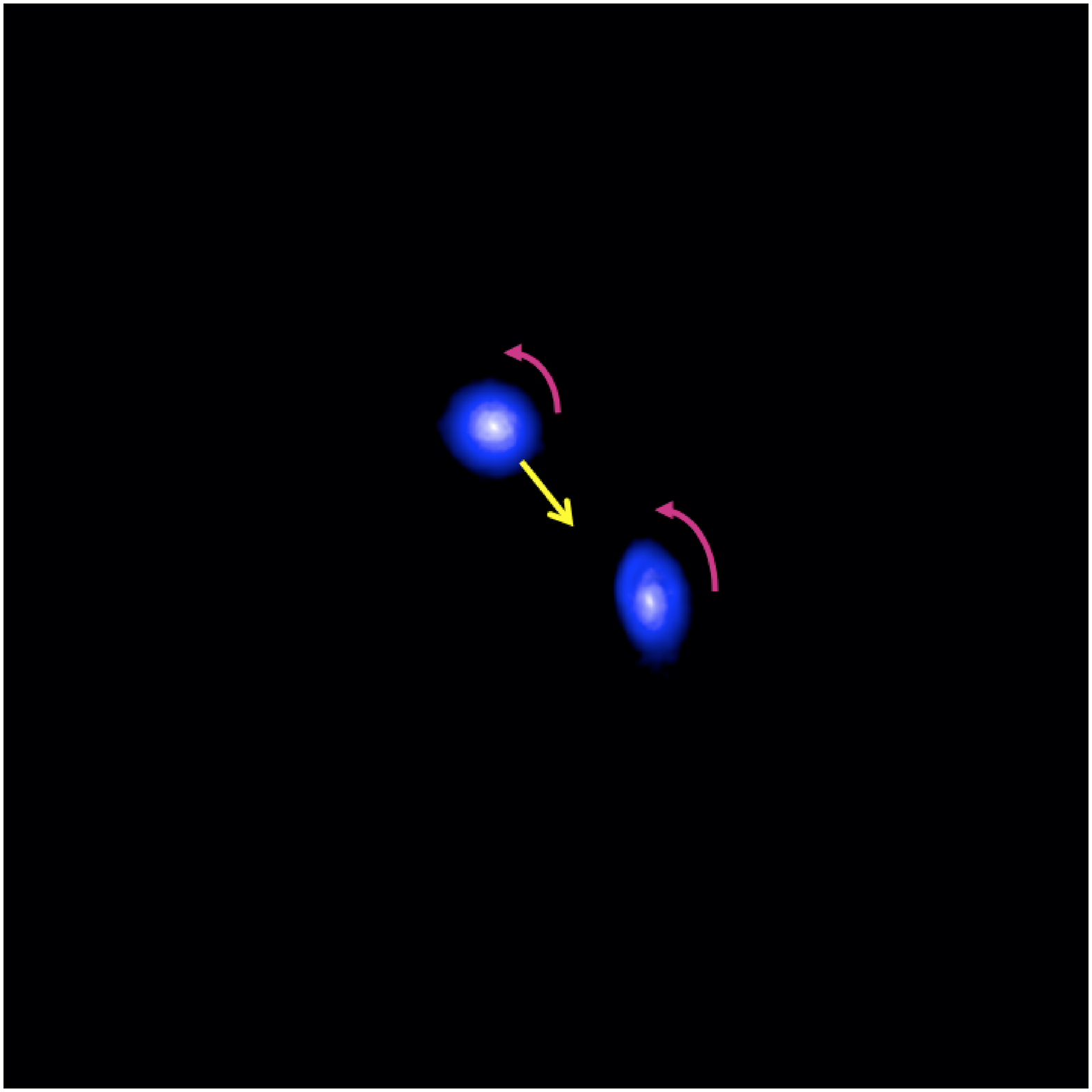}
    \includegraphics[width=3.cm]{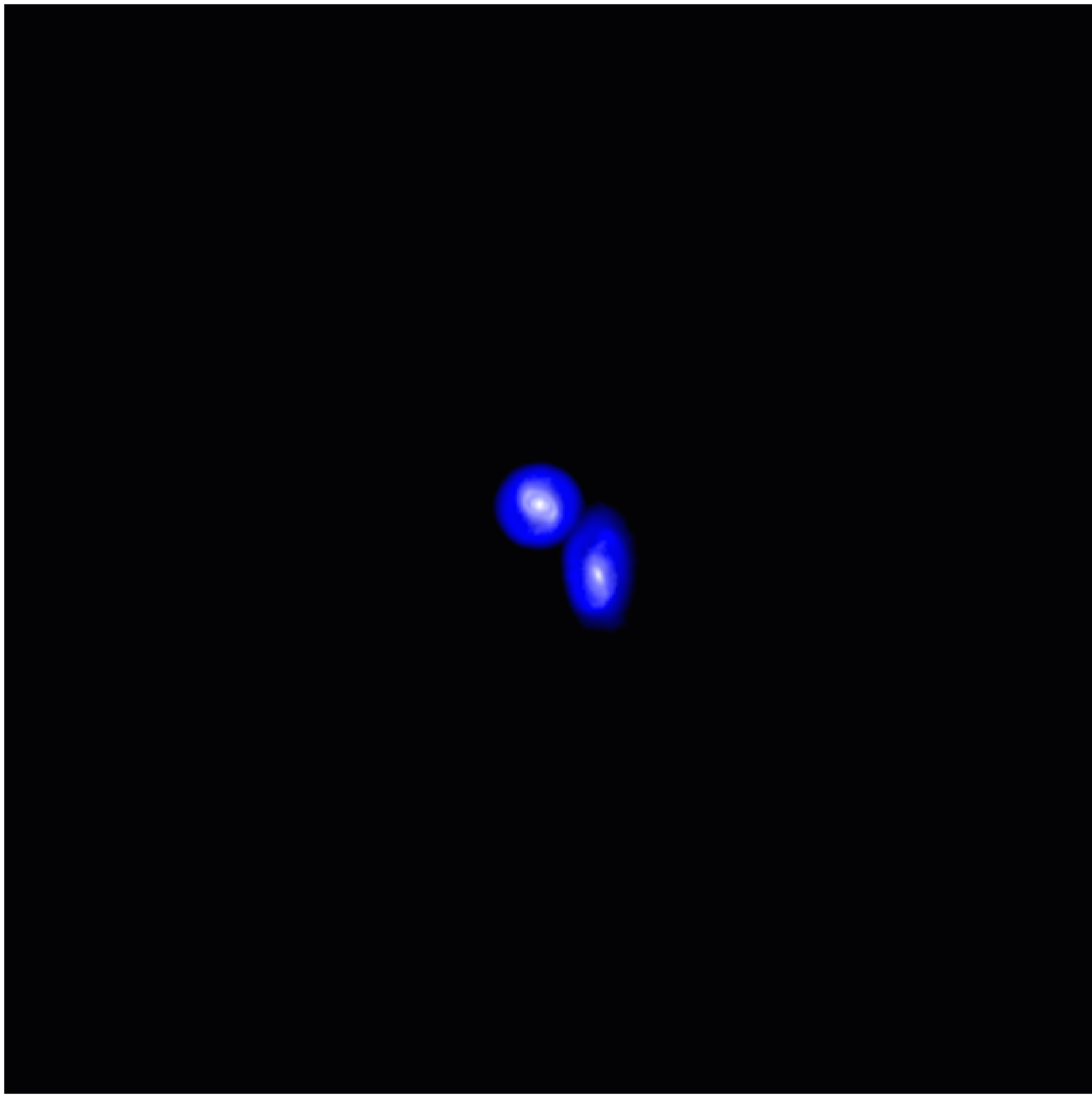}

    \includegraphics[width=3.cm]{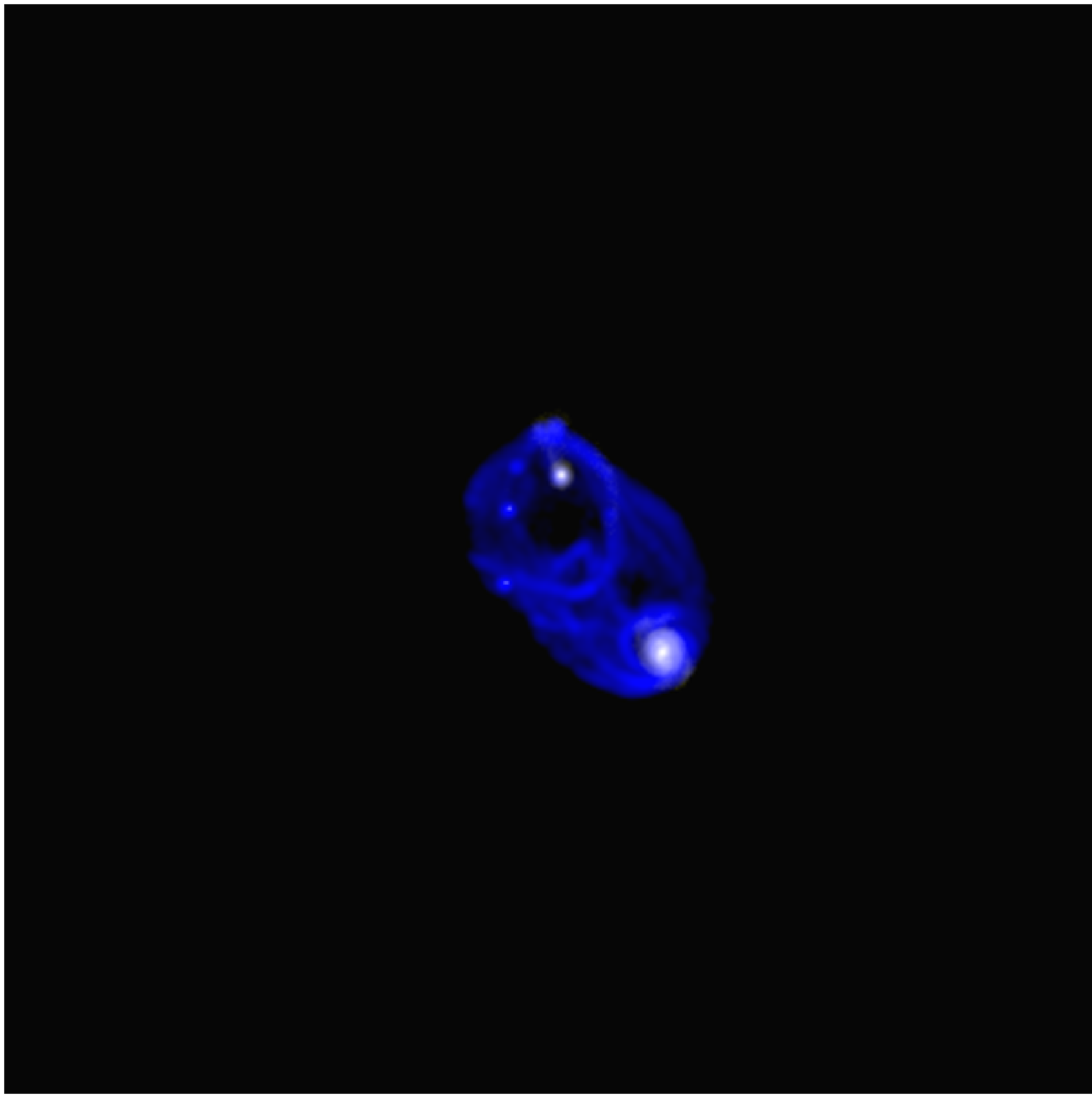}
    \includegraphics[width=3.cm]{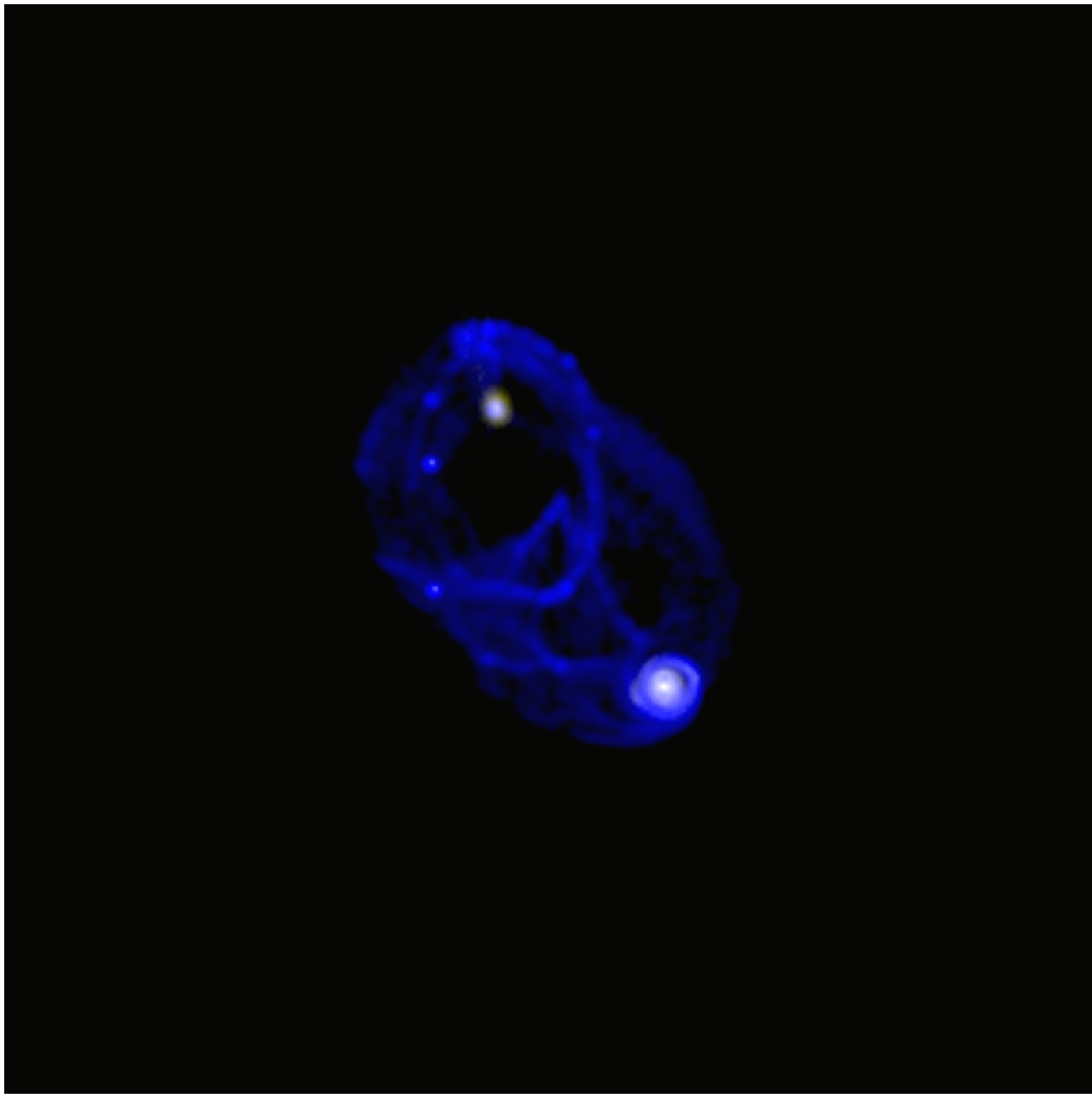}

    \includegraphics[width=3.cm]{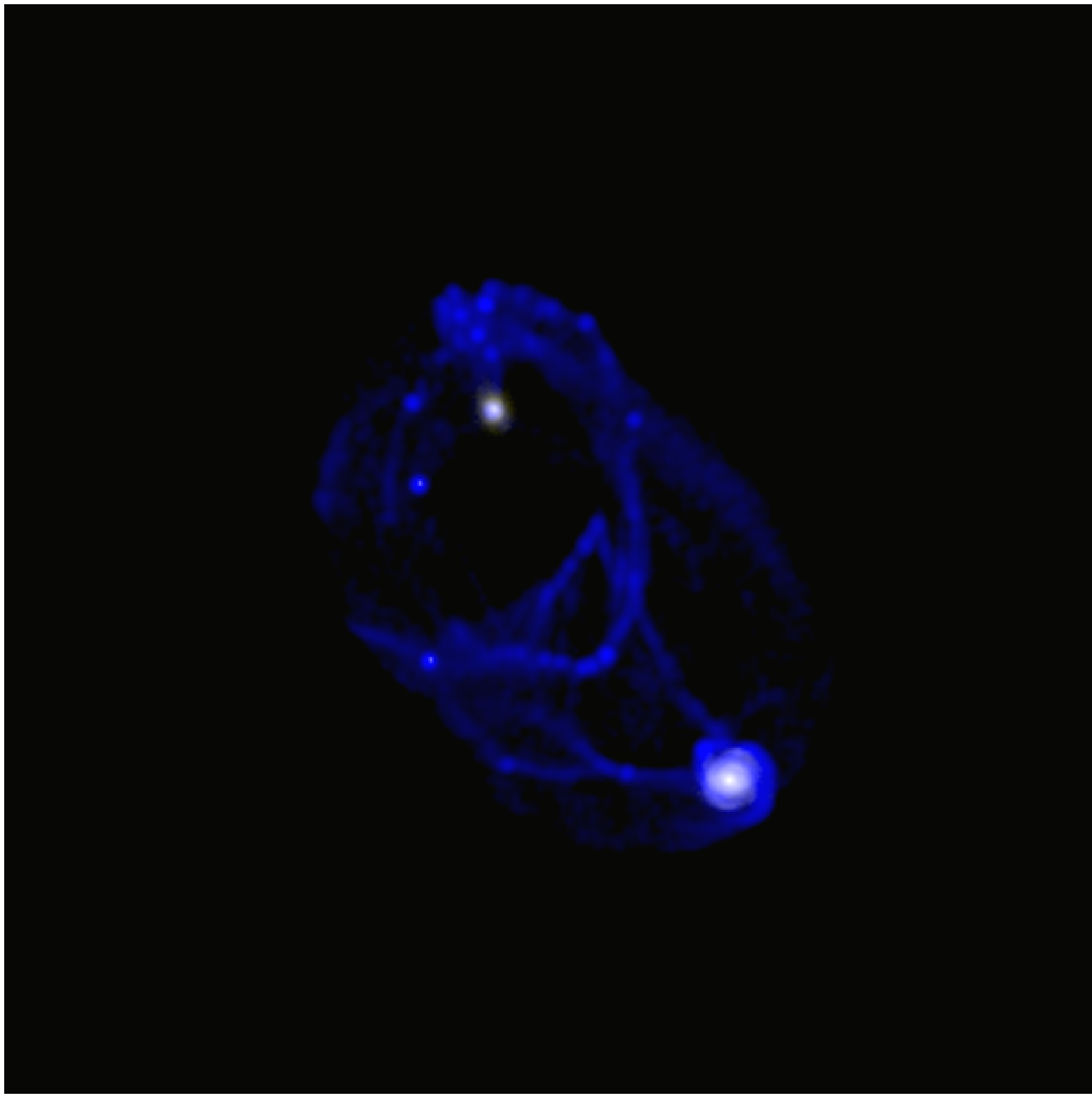}
    \includegraphics[width=3.cm]{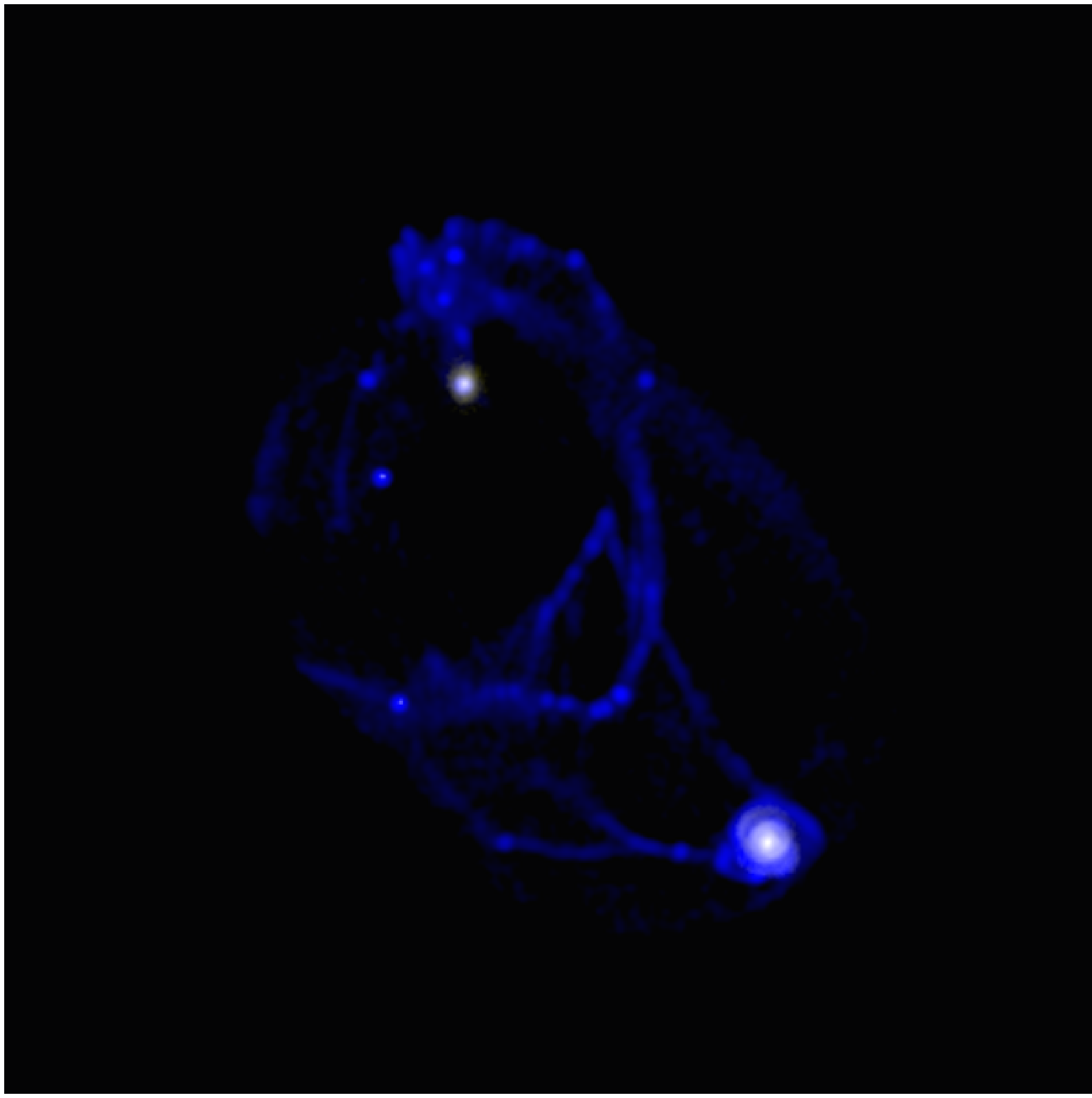}
  \end{minipage}
  \begin{minipage}{9.08cm}
    \includegraphics[width=9.08cm]{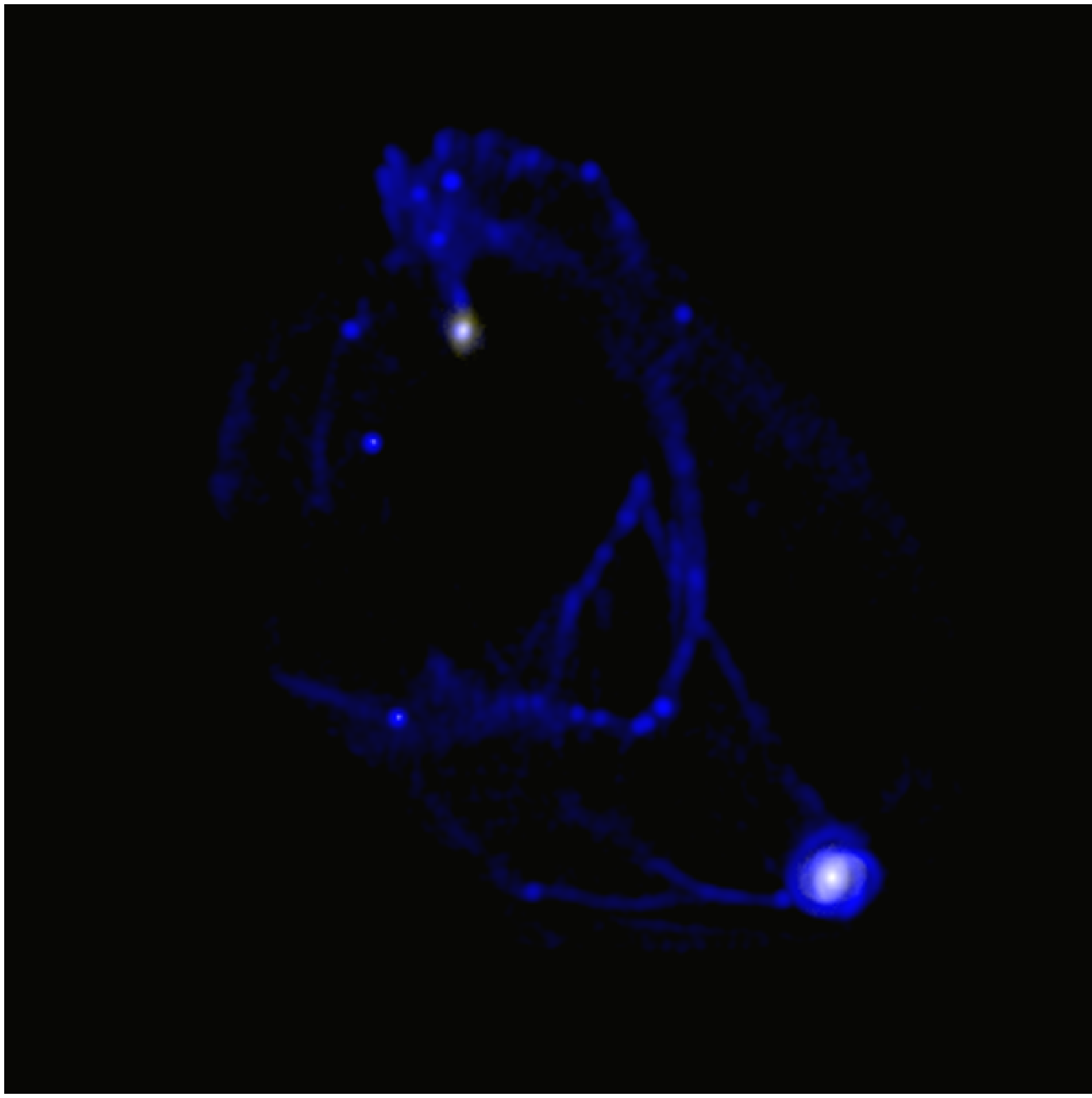}
  \end{minipage}

  \caption{Modeling the formation of the Leo ring in a head-on
    encounter between NGC 3384 and M96. Each panel measures 400 kpc on
    a side. The color coding traces both the gas density (in blue)
    and the stellar density (in yellow). The simulation is shown at
    six different instants in the six small panels to the left, with
    time intervals of 230~Myr.  For each snapshot the simulation is
    projected on the same way.  The top left frame shows the position
    of the two galaxies at the beginning of the simulation. In this
    projection, the interloper is initially in front of the target.
    The pink arrows indicate the rotation of each galaxy, whereas the
    yellow one indicates the initial velocity of the interloper.  The
    asymmetry in the disk of the target galaxy can be seen in the top
    middle panel. The right panel shows a snapshot $1.2$~Gyr after the
    collision.}
  \end{center}
  \label{fig:snapshots}
\end{figure*}

These results contradict the claim by \cite{Thilker09} that the gas
fueling the star-formation episode can only be very metal poor and
thus primordial. Then, the extended \hi\ structure probably results
from a past interaction in the group. In the following, we explore
this hypothesis with numerical simulations.

\section{Simulations}

We model a galaxy encounter using a particle-mesh sticky particle code
\citep[][and references therein]{Bournaud07}, with a softening length
of 150~pc.  The particle mass is $8 \times 10^4 M_{\odot}$ for gas,
$4 \times 10^5 M_{\odot}$ for stars, and $7.2 \times
10^5 M_{\odot}$ for dark matter (DM).  Gas dynamics is modeled through a
sticky particle scheme with $\beta_r = \beta_t = 0.45$ and star
formation is computed with a Schmidt--Kennicutt law
\citep{Kennicutt98a} with the star formation rate proportional to the
gas density to the exponent 1.4.

Among a set of 50 simulations of head-on collisions, we present here a
specific case that reproduces, at least qualitatively, the main
properties of the Leo ring. It is a relatively low-velocity encounter,
not because a high-velocity encounter is ruled out (high-velocity
encounters can form rings and tidal debris, as shown by
\citealt{Duc08,Bournaud07}), but because this configuration seems more
likely in the low velocity dispersion Leo group. Besides, initial
tests showed that the interloper had to be very massive to form a
giant ring.  In our model, NGC~3384 is the target disk of a head-on
collision with the massive galaxy M96.  M96 appeared to be the most
likely interloper, assuming that the \hi\ bridge linking it to the
ring was left over after the collision. Note that the key role played
by M96 in the formation of the ring had already been envisioned by
\cite{Rood85}. As for the target galaxy, the S0 galaxy NGC~3384 seemed
a better candidate than the other central galaxy (M105): in the
collision scenario, the target should initially be a gas-rich spiral,
and an interaction without merger is more likely to convert it into a
disky ETG (S0/Sa) than into a spherical elliptical such as M105
\citep{Bekki98,Bournaud03a}.

\begin{figure*}
  \begin{center}
  \includegraphics[width=8.76cm]{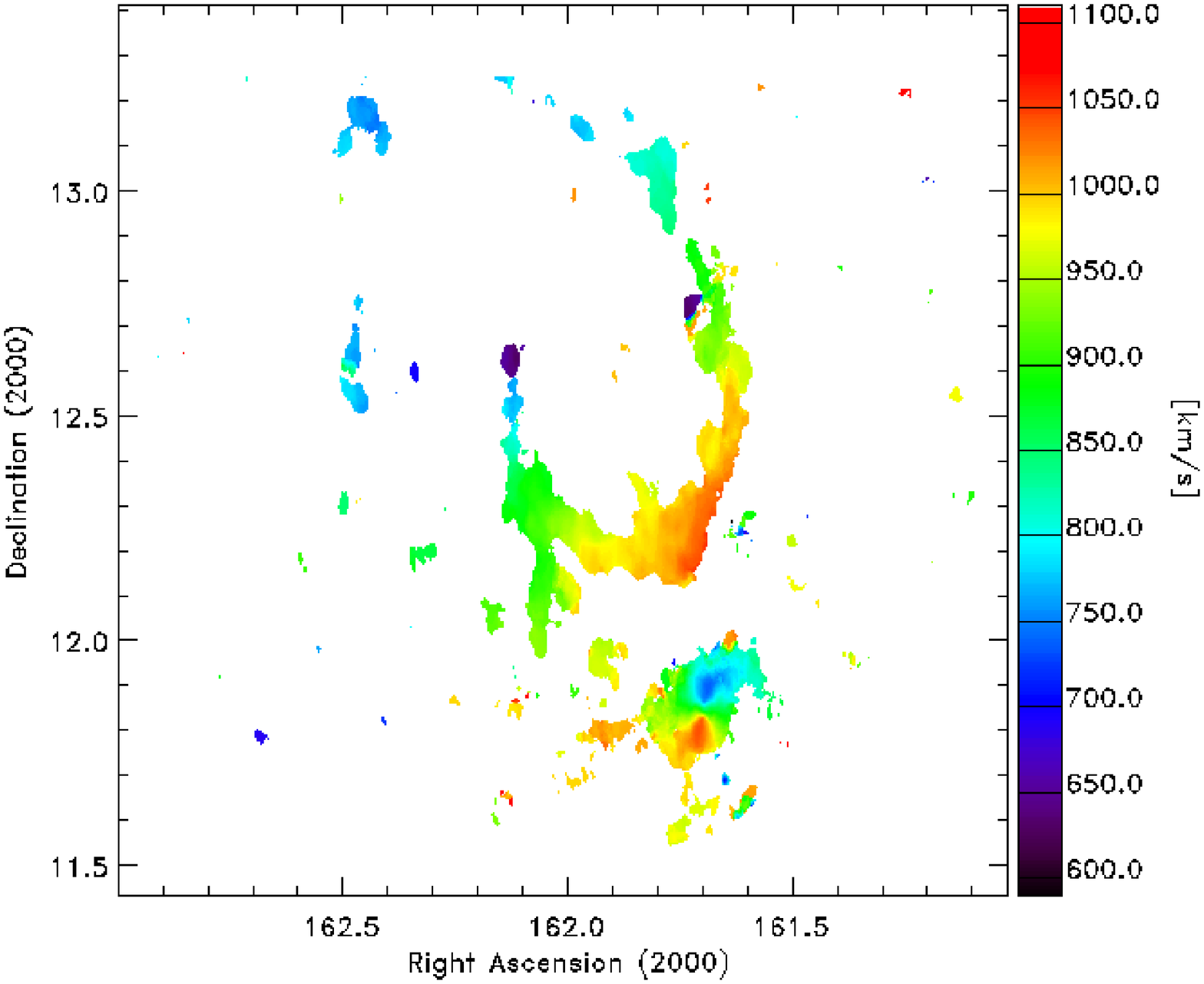}
  \includegraphics[width=8.76cm]{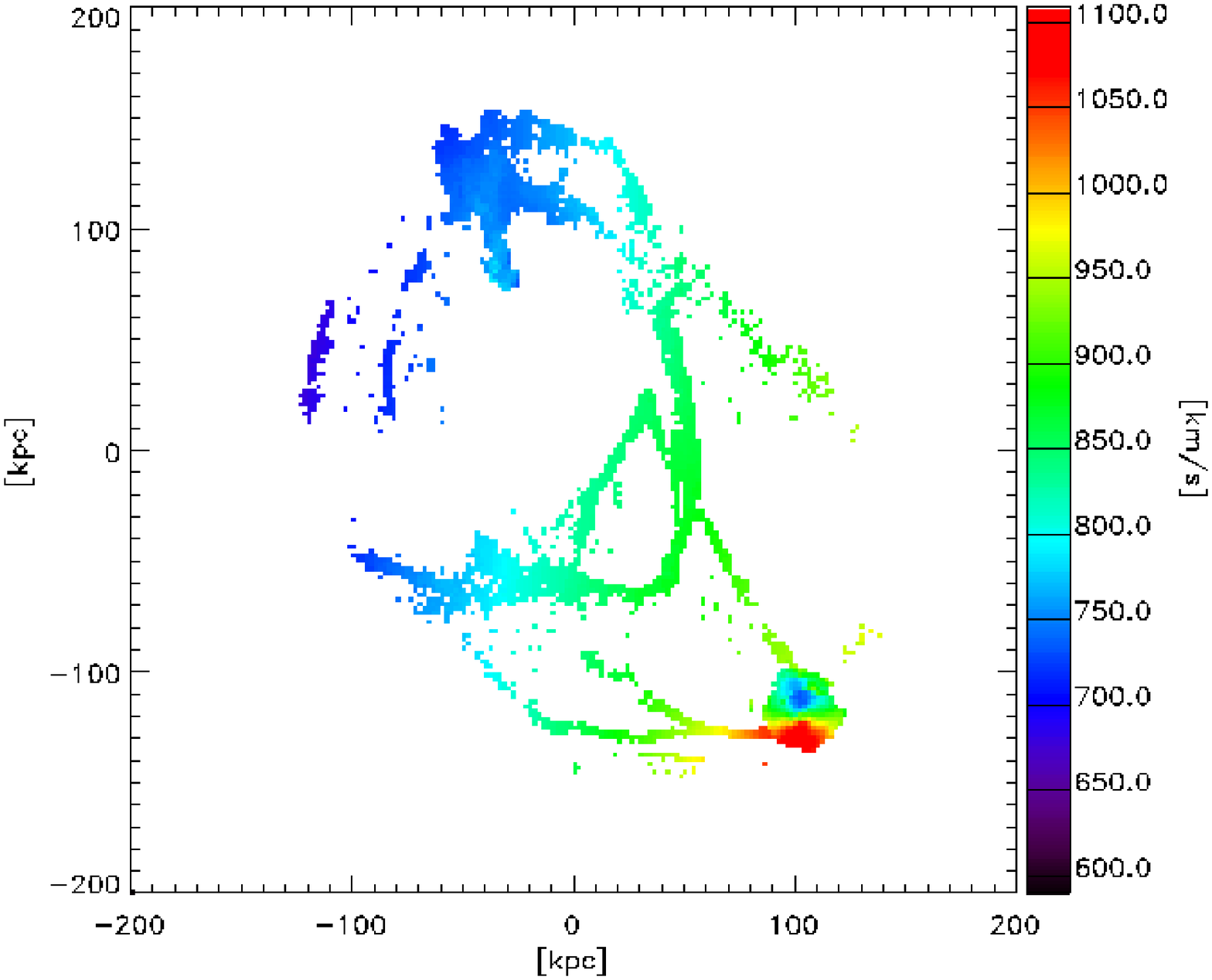}
  \caption{Left: velocity field of the \hi\ gas, as mapped by
    the WSRT (T. A. Oosterloo et al., in preparation). The field of view is $400
    \times 400$ kpc, assuming a distance of 11.6 Mpc. Right:
    velocity field of the gas in the numerical model of the
    collisional ring.}
  \end{center}
  \label{fig:velfield}
\end{figure*}

Both the interloper and the target galaxy are modeled with a disk of
gas and stars, a stellar bulge, and a DM halo. The target
galaxy is a gas-rich spiral galaxy with an extended gaseous disk. The
stellar disk has a mass $M_d = 1.92\times10^{10} M_{\odot}$
distributed according to a Toomre profile of scalelength $r_d =
3.375$~kpc.  The bulge is a Plummer sphere with a mass $M_b =
3.24\times10^{9} M_{\odot}$ and a characteristic radius $r_b =
1.5$~kpc. The gaseous disk is a homogeneous disk with a mass $M_g =
4.8\times10^{9} M_{\odot}$ truncated at $r_g = 18$~kpc. These
components are embedded in a spherical DM halo with
pseudo-isothermal profile, mass $M_h = 7.08\times10^{10} M_{\odot}$,
and core radius $r_h = 24.75$~kpc.

The interloper is a massive spiral galaxy resembling M96. Its
components have the following characteristics: $M_d =
5.04\times10^{10} M_{\odot}$, $r_d = 3.75$~kpc, $M_b =
1.2\times10^{10} M_{\odot}$, $r_b = 3$~kpc, $M_g =
5.64\times10^{9} M_{\odot}$, $r_g = 15$~kpc, $M_h =
1.98\times10^{11} M_{\odot}$, and $r_h = 15$~kpc.

We found that the strong asymmetry of the \hi\ structure cannot be
explained just by an off-center collision.  We then added an initial
asymmetry in the disk of the target by offsetting both the initial gas
disk center and the stellar components center by 8~kpc compared to the
DM halo center and relaxing the initial conditions in this
configuration, before simulating the interaction.  At the end of the
relaxation stage, the asymmetry (as defined in \citealp{Bournaud05})
is $\langle A_1 \rangle_{\mathrm{star}} \simeq 0.2$ and $\langle A_1
\rangle_{\mathrm{gas}} \simeq 0.4$.  This corresponds to a relatively typical
degree of asymmetry for a disk galaxy
\citep{Bournaud05}. \cite{Angiras06} found that one-third of group
galaxies have $\langle A_1 \rangle_{\mathrm{gas}} \geq 0.3$.

In the model presented in Figure~\ref{fig:snapshots}, the interloper
entered the system from the north at about 600 km~s$^{-1}$ and hit
the target with an impact parameter of 2.25~kpc. The two disks collide
perpendicularly and the orbit of the interloper has an inclination of
$45\degr$ with respect to the plane of the target disk.

The collision forms an expanding gas ring, mostly from NGC~3384
material. 1.2~Gyr later, the model reproduces, at least qualitatively,
the main observational features: the size, mass, column density, and
asymmetry of the \hi\ ring; the bridge of \hi\ linking the ring to
M96; the projected systemic velocity of the two colliding galaxies,
and the global kinematics of the gas along the ring and in M96, as
shown in Figures~\ref{fig:snapshots} and \ref{fig:velfield}.

The projected position of the target galaxy relative to the ring and
the mass in the northern end of the ring are not exactly
reproduced. There is also some mismatch between the gas velocities in
the interloper and the ring. Nevertheless, the main intriguing
properties of the Leo ring are reproduced, such as the strong
asymmetry even if the exact orientation of this asymmetry is not just
like in the Leo ring itself.  Our simulation should be seen as a proof
of concept that relatively common collisions with moderate velocities
between two galaxies typical for the Leo group can easily form a ``Leo
ring-like'' system. Matching all detailed properties accurately is not
our purpose and would induce too many free parameters: the tidal
field from other nearby galaxies (such as M105), and even of the whole
Leo group, would have to be accounted for since large-scale fields can
affect the result of a pair interaction \citep{Martig08}.

\section{Discussion and conclusions}
\label{sec:disc}

The origin of the Leo ring has been actively debated for more than two
decades. If made of primordial gas, it would be the only such giant
\hi\ structure known in the nearby universe. The absence of an optical
counterpart was often suggested to preclude a collisional origin. Our
deep optical images have confirmed that indeed it is not associated with
a diffuse stellar component brighter than 28 mag~arcsec$^{-2}$.
However, as shown by our numerical model of the Leo ring and previous
simulations \citep{Bournaud07}, collisional rings are not expected to
systematically contain a significant populations of old stars expelled
from the progenitor galaxy. Old stars can spread in a halo rather than
along the ring, as is the case in our model with a predicted
brightness fainter than $\sim$ 29~mag~arcsec$^{-2}$.

The Leo ring is striking for not having managed to convert a
significant fraction of its gas into stars. If primordial, this would
require unexampled long-term stability conditions \citep[stabilization
  by a deep potential such as the Leo group could nevertheless quench
  star formation;][]{martig09}. Our imaging program has revealed the
presence of compact, optical counterparts to the far-ultraviolet
sources detected by \textit{GALEX}. These are likely young stellar associations
formed in situ in the \hi\ ring, with very low star formation rates
that are consistent with the low \hi\ surface densities.
\cite{Thilker09} had estimated a very low metallicity from UV data,
suggesting the presence of primordial gas. Adding three data points in
the optical bands to the SED of one of the UV clumps, we find it most
consistent with that of a star-forming object in the NGC~5291 ring,
known to be pre-enriched and of collisional origin \citep{Bournaud07}.
Thus, available data are actually consistent with the Leo ring being
made-up of pre-enriched material.  Obtaining a direct measurement of
its metallicity would give the final word.

If the Leo ring is not made-up of primordial gas, its formation
should involve pre-enriched gas expelled during galaxy--galaxy or
galaxy--group interactions. Ram pressure would require a very dense
intracluster medium to produce such a massive and extended structure,
a condition not met in a relatively loose group like Leo. Note also
that ram pressure usually forms one-sided tails attached to the parent
galaxy, not ring-like structures \citep{tonnessen09}. The tidal
potential of the group could also affect the gas content of galaxies,
but is generally a second-order effect compared to direct collisions
in galaxy pairs.

A more likely hypothesis is that the Leo ring has the same origin as
the confirmed collisional ring around NGC~5291
\citep{Longmore79,Bournaud07} or the famous Cartwheel ring
\citep{Horellou01} --- the later being much denser and actively
star forming.  Our numerical model reproduces the main characteristics
of the Leo ring, after a galaxy collision at a relatively moderate
speed, quite plausible in the Leo group. The proposed progenitor
galaxy, NGC 3384, does not show strong morphological disturbances,
even in our deep imaging data. This is however expected in our model,
that produces a gaseous collisional ring, an extended and very faint
(unobservable) stellar halo, but leaves little signs of collision such
as classical tails and shells around the parent
galaxy. \texttt{SAURON} observations of NGC~3384
\citep{Emsellem04,Kuntschner06} revealed a barred rotating disk, with
a low gas and dust content and without very recent star formation,
which is also broadly consistent with our model where the pre-existing
spiral is converted into a fastly rotating ETG by the 1.2 Gyr old
collision.

Large collisional rings can form in groups of galaxies and take an
appearance a priori suggesting primordial gas rather than tidal
debris. The Leo ring appears fully consistent with being the result of
such a process.


\acknowledgments

We acknowledge HPC resources from GENCI-CCRT (grant 2009-042192),
support from Agence Nationale de la Recherche (ANR-08-BLAN-0274-01),
and \textit{GALEX} archival data. We thank Fran\c{c}oise Combes for valuable
discussions.

{}

\end{document}